# Superconducting nanodrop model

Iogann Tolbatov

*Physics and Engineering Department, Kuban State University, Krasnodar, Russia*

(talbot1038@mail.ru)

Superconducting nanodrop model is constructed on the basis of the assumption of equality of the free energy functional gradient term to zero due to the order parameter value constancy in the superconducting nanodrop volume. In the investigation, we have defined the superconducting nanodrop thermodynamical characteristics: partition function, the magnitudes of the jumps in entropy and specific heat at the second order phase transition, total energy, energy fluctuation, specific heat, free energy, and entropy.

Fluctuation theory is that intermediate link that connects many disparate theoretical models and experimental superconductivity physics data as one of the basic methods of study of both the traditional low-temperature and the high-temperature superconducting structures. Recently, researchers have paid attention to the superconducting nanostructures [1-5], the study of which is a necessary springboard for the development of nanotechnology as a foundation for scientific and technological revolution of the XXI century.

The subject of our research is fluctuations in superconducting nanostructures. Using the methodology of the theory of fluctuations, we have solved the problem of the superconducting nanodrop model construction. The study was conducted within the Ginzburg – Landau formalism frameworks. According to the hypothesis, the gradient term of the free energy functional is equal to zero due to the supposed order parameter value constancy in the superconducting nanodrop volume.

The theoretical significance of the study is in the superconducting nanodrop model construction and in its results comprehension from the standpoint of fundamental physics. In the constructed model, we have defined the thermodynamic characteristics of superconducting nanodrop. These include the partition function of the superconducting nanodrop system (the identification of this quantity means the model construction), the magnitudes of the jumps in entropy and specific heat at the second order phase transition, total energy, energy fluctuation, specific heat, free energy, and entropy. The significance of the study lies in the fact that the study of the superconducting properties of nanodrops sheds light on the characteristics of superconducting nanogranules comprising the superconducting nanodrops, and the properties of granular superconductors, which, in their turn, are structures constructed from superconducting nanodrops.

**Investigation apparatus**

As it is known from statistical physics, a complete description of the thermodynamic properties of the system can be obtained by calculating its partition function. We use the system of units $\hbar = k_B = c = 1$:

$$Z = tr\left\{\exp\left(-\frac{\hat{H}}{T}\right)\right\}. \tag{1}$$

In the superconducting transition vicinity, the fluctuation Cooper pairs of bosonic nature appear along with the fermionic electron states in the metal. They can be described by classical boson fields $\Psi(\vec{r})$, which may be regarded as the fluctuation Cooper pairs wave functions. Thus, the trace calculation in (1) can be divided into the summation over "fast" electron degrees of freedom and the subsequent functional integration over all possible configurations of the "slow" wave functions of the fluctuation Cooper pairs:

$$Z = \int D^2\Psi(\vec{r})\,\Xi[\Psi(\vec{r})], \tag{2}$$

where

$$\Xi[\Psi(\vec{r})] = \exp\left(-\frac{F[\Psi(\vec{r})]}{T}\right) \tag{3}$$

is the system partition function summed over electronic degrees of freedom with a fixed boson field $\Psi(\vec{r})$, where $F[\Psi(\vec{r})]$ is the Ginzburg – Landau (GL) functional.

The GL functional can be normalized in two ways [6].

**Normalization, where the coefficient $m$ is equal to the mass of the electron.**
The GL functional is written in the form:

$$F[\Psi(\vec{r})] = F_N + \int dV \left\{ a|\Psi(\vec{r})|^2 + \frac{b}{2}|\Psi(\vec{r})|^4 + \frac{1}{4m}|\nabla\Psi(\vec{r})|^2 \right\}. \tag{4}$$

Let us discuss its coefficients. In accordance with the hypothesis of Landau, the coefficient $a$ vanishes at the transition point and depends linearly on $T-T_C$. It is convenient to separate the dependence on reduced temperature $\varsigma$ characterizing degree of the system closeness to the transition point and to represent $a$ as $a = \alpha T_C \varsigma$. All coefficients $\alpha$, $b$ and $m$ are assumed to be positive and independent of temperature. In the phenomenological theory of Ginzburg – Landau, the order parameter $\Psi$ normalization is usually chosen in such a way that the coefficient $m$ corresponds to the mass of the electron. The coefficient $\alpha$ for $D$-dimensional pure superconductor is given by

$$\alpha_{(D)} = \frac{2D\pi^2}{7\zeta(3)} \frac{T_C}{E_F}. \tag{5}$$

**Normalization, at which the order parameter and the gap in the spectrum of superconductor single-particle excitations are equal.** It is more convenient to use another normalization, where the order parameter denoted as $\Delta(\vec{r})$ in the homogeneous case is equal to the gap in the spectrum of superconductor single-particle excitations. Close to $T_C$, the microscopic theory allows us to define the superconductor free energy in the form of the GL expansion over the parameter $\Delta(\vec{r})$ powers:

$$F[\Delta(\vec{r})] = \int \left[ A\Delta^2 + \frac{B}{2}\Delta^4 + C_{(D)}(\nabla\Delta)^2 \right] dV, \tag{6}$$

where

$$A = \upsilon \ln \frac{T}{T_C}, \tag{7}$$

$$B = \frac{7\zeta(3)}{8\pi^2 T^2} \upsilon. \tag{8}$$

Coefficient $C_{(D)}$ is related to the square of the coherence length:

$$C_{(D)} = \upsilon \xi_{(D)}^2 (T\tau). \qquad (9)$$

**Study hypothesis**

We consider the superconducting system with structural disorder, which causes the local critical temperature smooth variation in space. It can be described by the GL functional with coefficients being some random functions of coordinates. We suppose that $T > T_C$, i. e., $\bar{a} > 0$.

If at some sufficiently large area, such a situation $a = \bar{a} + \delta a(\vec{r}) < 0$ is realized, then the superconducting drop is possible to appear in it. "Sufficiently large" means that the region should be so large that the effect of proximity is unable to suppress the superconductivity arising in it. That is why, the probability of such areas formation is low. The problem of calculating this probability (i. e., finding the density of superconducting drops) is a special case. Assuming the characteristic scale of the structural disorder manifestations less than $\xi$, we can conclude that the distribution functions of random variables interesting to us have the Gaussian form. For example, for $a = \bar{a} + \delta a(\vec{r})$:

$$P[\delta a(\vec{r})] = C \exp\left[-\frac{1}{W} \int [\delta a(\vec{r})]^2 d\vec{r}\right], \qquad (10)$$

where $W$ is a phenomenological parameter, which can be determined experimentally by measuring the critical current at $T < T_C$.

The term "superconducting drop" should be understood as the region of space, in which the GL factor is $a < 0$, and this region should be large with a such characteristic size $L$ that the fluctuations of the order parameter modulus $|\Delta|$ could be neglected ($T = 0$ means that the superconducting gap $\Delta$ in it would exceed the corresponding average distance between the levels of size quantization $\delta$). As a result, from the GL equation with the coefficient $a(\vec{r})$, which is a function of $\vec{r}$, we can find the quantity $|\Psi|$, and then calculate the density of such drops, i. e., the probability of finding it at the point $\vec{r}$. We will solve the problem qualitatively. Due to the proximity effect, the order parameter within the drop size $L$ can be estimated as [7]:

$$|\Psi|^2 = \frac{1}{b}\left(-a - \frac{1}{mL^2}\right), \qquad (11)$$

The order parameter $\Delta$ value in the simplest version of the Bardeen – Cooper – Schrieffer (BCS) theory is valid for homogeneous superconductor in the absence of magnetic field and paramagnetic impurities, it coincides with the size of the gap in the quasiparticles spectrum. Order parameter $\Psi$ phenomenologically introduced may be associated with the corresponding parameter $\Delta$ of the microscopic theory through the relation:

$$\Psi = \sqrt{4mC_{(D)}}\Delta. \qquad (12)$$

With this definition, the microscopic theory allows us to define exact values for the phenomenological coefficients $\alpha$ and $b$:

$$4m\alpha T_C = \xi^{-2}; \quad \frac{\alpha^2}{b} = \frac{8\pi^2}{7\zeta(3)}\upsilon, \qquad (13)$$

where $\zeta(x)$ is the Riemann zeta function, $\zeta(3) = 1{,}202$.

In the phenomenological theory frameworks, $m$ is usually identified with the mass of free electron in such a way that the mass of the Cooper pair is equal to intuitively expected $2m$ (4). With this choice of mass, parameter $\alpha$ for pure $D$-dimensional superconductor is equal to:

$$\alpha_{(D)} = \frac{2D\pi^2}{7\zeta(3)}\frac{T_C}{E_F}. \qquad (14)$$

We consider a homogeneous superconducting nanodrop in the absence of magnetic field and paramagnetic impurities. Consequently, the gradient term in the expression for the GL functional (6) is zero: $(\nabla\Delta) = 0$. The order parameter defined by the functional (6) is constant in the drop volume. The GL functional now takes the form:

$$F[\Delta(\vec{r})] = \int\left[A\Delta^2 + \frac{B}{2}\Delta^4\right]dV. \qquad (15)$$

**Superconducting nanodrop model construction**

The superconducting nanodrop volume is:

$$V = \frac{\pi L^3}{6}. \qquad (16)$$

From (11), (12), it follows that the the order parameter square $\Delta^2$ is defined by the expression:

$$\Delta^2 = \frac{\left(-a - \frac{1}{mL^2}\right)}{4mC_{(D)}b}. \tag{17}$$

From (7), (8), (9) and (13), it follows that coefficients $a$ and $b$ (11) are equal respectively to:

$$a = \frac{\varsigma}{4m\xi^2}, \tag{18}$$

$$b = \frac{7\zeta(3)}{128\pi^2 \upsilon m^2 T_C^2 \xi^4}. \tag{19}$$

Hence, the order parameter square $\Delta^2$ (17) equals:

$$\Delta^2 = \frac{32\pi^2 T_C^2 \xi^2 \left(-\frac{\varsigma}{4\xi^2} - \frac{1}{L^2}\right)}{7\zeta(3)T\tau}. \tag{20}$$

Due to the studied object homogeneity, we rewrite (15) in the form:

$$F[\Delta(\vec{r})] = A\Delta^2 V + \frac{B}{2}\Delta^4 V, \tag{21}$$

where the superconducting drop volume $V$ is defined by expression (16).

From formulae (7), (8), (18), (19) and (20), we can find that in (21), $A\Delta^2 V$ and $\frac{B}{2}\Delta^4 V$ can be expressed:

$$A\Delta^2 V = \frac{-4\pi^3 \upsilon L T_C^2 \varepsilon}{21\zeta(3)T\tau}\left(\varepsilon L^2 + 4\xi^2\right), \tag{22}$$

$$\frac{B}{2}\Delta^4 V = \frac{2\pi^3 \upsilon T_C^4}{21\zeta(3)T^4 \tau^2 L}\left(\varepsilon^2 L^4 + 16\xi^4 + 8\varepsilon\xi^2 L^2\right). \tag{23}$$

As we calculate the free energy functional (21), we reject the terms, which include $\xi^2$ because of their smallness. Thus, the functional is simplified to the following:

$$F = \frac{2\pi^3 \upsilon T_C^2 \varepsilon^2 L^3}{21\zeta(3)\tau^2 T^4}\left(T_C^2 - 2\tau T^3\right). \tag{24}$$

The system partition function summed over electronic degrees of freedom for a fixed bosonic field $\Psi(\vec{r})$ (3) is written as:

$$\Xi[\Psi(\vec{r})] = \exp\left(-\frac{2\pi^3 \upsilon T_C^2 \varepsilon^2 L^3}{21\zeta(3)\tau^2 T^5}\left(T_C^2 - 2\tau T^3\right)\right). \tag{25}$$

Integration (2) in the superconducting nanodrop system partition function is reduced to multiplication by factor $\frac{4mC_{(D)}}{2}\Delta^2$ due to the order parameter constancy at each point of nanodrop:

$$Z = \frac{-16\pi^2 m\upsilon T_C^2 \xi^2}{7\zeta(3)L^2}\left(\varepsilon L^2 + 4\xi^2\right)\exp\left(-\frac{2\pi^3 \upsilon T_C^2 \varepsilon^2 L^3}{21\zeta(3)\tau^2 T^5}\left(T_C^2 - 2\tau T^3\right)\right). \tag{26}$$

Due to the apparent smallness of $\xi^4$, equation (26) can be rewritten as follows:

$$Z = \frac{-16\pi^2 m\upsilon T_C^2 \xi^2 \varepsilon}{7\zeta(3)}\exp\left(-\frac{2\pi^3 \upsilon T_C^2 \varepsilon^2 L^3}{21\zeta(3)\tau^2 T^5}\left(T_C^2 - 2\tau T^3\right)\right). \tag{27}$$

The superconducting nanodrop model is constructed.

**Determination of superconducting nanodrop thermodynamic characteristics**

Let us rewrite (27) in the form:

$$Z = qT_C^2 \varepsilon \exp\left(\frac{y(T - T_C)^2(T_C^2 - 2\tau T^3)}{T^5}\right), \tag{28}$$

where $q = \frac{-16\pi^2 m\upsilon \xi^2}{7\zeta(3)}$ and $y = -\frac{2\pi^3 \upsilon L^3}{21\zeta(3)\tau^2}$. \tag{29}

If the partition function is expressed by means of the reduced temperature $\varsigma$, equation (28) takes the form:

$$Z = qT_C^2 \varsigma \exp\left(\frac{y\varsigma^2}{T_C(1+\varsigma)^5}\right). \tag{30}$$

If the partition function is expressed by means of the inverse temperature $\beta = \frac{1}{T}$, equation (28) takes the form:

$$Z = q\left(\frac{1}{\beta} - T_C\right)^2 \exp\left[y\beta^5\left(\frac{1}{\beta} - T_C\right)^2\left(T_C^2 - \frac{2\tau}{\beta^3}\right)\right]. \tag{31}$$

We take into account also that

$$\frac{\partial Z}{\partial \varsigma} = qT_C^2 \exp\left(\frac{y\varsigma^2}{T_C(1+\varsigma)^5}\right)\left[1 + \frac{y\varsigma^2(2-3\varsigma)}{T_C(1+\varsigma)^6}\right], \qquad (32)$$

$$\frac{\partial^2 Z}{\partial \varsigma^2} = qT_C^2 \exp\left(\frac{y\varsigma^2}{T_C(1+\varsigma)^5}\right)\left[\left(1 + \frac{y\varsigma^2(2-3\varsigma)}{T_C(1+\varsigma)^6}\right)\frac{y\varsigma(2-3\varsigma)}{T_C(1+\varsigma)^6} + \frac{y\varsigma(4+7\varsigma-27\varsigma^2)}{T_C(1+\varsigma)^7}\right]. \qquad (33)$$

**Magnitude of the jump in entropy at second order phase transition in superconducting nanodrop:**

$$\Delta S = -Z^{-1}\frac{\partial Z}{\partial \varsigma}. \qquad (34)$$

Substituting expressions (30) and (32) in (34), we obtain the value of the jump in entropy:

$$\Delta S = \frac{1}{\varsigma} + \frac{y\varsigma(2-3\varsigma)}{T_C(1+\varsigma)^6}. \qquad (35)$$

**Magnitude of the jump in the specific heat at second order phase transition in superconducting nanodrop:**

$$\Delta C = Z^{-1}\frac{\partial^2 Z}{\partial \varsigma^2} - Z^{-2}\left(\frac{\partial Z}{\partial \varsigma}\right)^2. \qquad (36)$$

Recalling that $\varsigma \ll 1$, we obtain:

$$\Delta C = \frac{4y - 2yT_C - T_C}{T_C(1+\varsigma)^{12}}. \qquad (37)$$

**Total energy of the system:**

$$\langle E \rangle = -\frac{\partial \ln Z}{\partial \beta}. \qquad (38)$$

Taking into account (31) and (32), we obtain the total energy of the superconducting nanodrop from (38):

$$\langle E \rangle = 2 - (4y\tau T_C)\beta + (8y\tau T_C^2)\beta^2 - (4y\tau T_C^3 + 3yT_C^2)\beta^3 + (11yT_C^3)\beta^4 -$$
$$- (13yT_C^4)\beta^5 + (5yT_C^5)\beta^6 \qquad (39)$$

**Superconducting nanodrop energy fluctuation is defined by the formula:**

$$\langle \delta E^2 \rangle = \langle (E - \langle E \rangle)^2 \rangle = \frac{\partial^2 \ln Z}{\partial \beta^2}. \qquad (40)$$

Substituting (31) and (33) in (40), we obtain:

$$\langle \delta E^2 \rangle = 4y\tau T_C - (16y\tau T_C^2)\beta + (12y\tau T_C^3 + 9yT_C^2)\beta^2 - (44yT_C^3)\beta^3 +$$
$$+ (65yT_C^4)\beta^4 - (30yT_C^5)\beta^5 \qquad (41)$$

**Superconducting nanodrop specific heat is defined by the formula:**

$$C_v = \frac{\partial \langle E \rangle}{\partial T} = \frac{\langle \delta E^2 \rangle}{T^2}. \qquad (42)$$

Thus, substituting in (42) expression (41), we obtain:

$$C_v = 4y\tau T_C \beta^2 - (16y\tau T_C^2)\beta^3 + (12y\tau T_C^3 + 9yT_C^2)\beta^4 - (44yT_C^3)\beta^5 +$$
$$+ (65yT_C^4)\beta^6 - (30yT_C^5)\beta^7 \qquad (43)$$

**Superconducting nanodrop free energy is defined by the expression:**

$$F = -T \ln Z. \qquad (44)$$

Consequently, from expressions (31) and (44), we find:

$$F = -\frac{2}{\beta} \ln\left[\sqrt{q}\left(\frac{1}{\beta} - T_C\right)\right] - y\beta^4 \left(\frac{1}{\beta} - T_C\right)^2 \left(T_C^2 - \frac{2\tau}{\beta^3}\right), \qquad (45)$$

**Superconducting nanodrop entropy is defined by the formula:**

$$S = -\frac{\partial F}{\partial T}. \qquad (46)$$

Thus, from expressions (45) and (46), we obtain:

$$S = 2\ln\left[\sqrt{q}\left(\frac{1}{\beta} - T_C\right)\right] + \frac{2}{(1-\beta T_C)} + 4yT_C^4\beta^5 - 6yT_C^3\beta^4 + 2yT_C^2\beta^3 - 2y\tau T_C^2\beta^2 + 2y\tau, \qquad (47)$$

**Conclusions**

1. Superconducting nanodrop model is constructed on the basis of the stated hypothesis of equality of the free energy functional gradient term to zero due to the order parameter value constancy in the superconducting nanodrop volume and the order parameter evaluation [8] within the nanodrop size $L$ limits.

2. In the investigation, we have defined the superconducting nanodrop thermodynamical characteristics: partition function, the magnitudes of the jumps in entropy and specific heat at the second order phase transition, total energy, energy fluctuation, specific heat, free energy, and entropy.